\documentclass[twocolumn]{pasj00}
\usepackage[dvips]{color}

\begin{document}
\SetRunningHead{N. Narita et al.}{A Common Proper
Motion Stellar Companion to HAT-P-7}
\Received{2012/08/31}%{yyyy/mm/dd}
\Accepted{2012/09/20}%{yyyy/mm/dd}

\title{A Common Proper Motion Stellar Companion to HAT-P-7}
%%% begin:list of authors

\author{
Norio \textsc{Narita},\altaffilmark{1,25}
Yasuhiro H. \textsc{Takahashi},\altaffilmark{1,2} 
Masayuki \textsc{Kuzuhara},\altaffilmark{1,3}
Teruyuki \textsc{Hirano},\altaffilmark{4}
Takuya \textsc{Suenaga},\altaffilmark{5}\\
Ryo \textsc{Kandori},\altaffilmark{1}
Tomoyuki \textsc{Kudo},\altaffilmark{6}
Bun'ei \textsc{Sato},\altaffilmark{7}
Ryuji \textsc{Suzuki},\altaffilmark{1}
Shigeru \textsc{Ida},\altaffilmark{7}
Makiko \textsc{Nagasawa},\altaffilmark{7}\\
Lyu \textsc{Abe},\altaffilmark{8}
Wolfgang \textsc{Brandner},\altaffilmark{9}
Timothy D. \textsc{Brandt},\altaffilmark{10}
Joseph \textsc{Carson},\altaffilmark{11}
Sebastian E. \textsc{Egner},\altaffilmark{6}\\
Markus \textsc{Feldt},\altaffilmark{9}
Miwa \textsc{Goto},\altaffilmark{12}
Carol A. \textsc{Grady},\altaffilmark{13}
Olivier \textsc{Guyon},\altaffilmark{6}
Jun \textsc{Hashimoto},\altaffilmark{1}\\
Yutaka \textsc{Hayano},\altaffilmark{6}
Masahiko \textsc{Hayashi},\altaffilmark{1}
Saeko S. \textsc{Hayashi},\altaffilmark{6}
Thomas \textsc{Henning},\altaffilmark{9}
Klaus W. \textsc{Hodapp},\altaffilmark{14}\\
Miki \textsc{Ishii},\altaffilmark{6}
Masanori \textsc{Iye},\altaffilmark{1}
Markus \textsc{Janson},\altaffilmark{10}
Gillian R. \textsc{Knapp},\altaffilmark{10}
Nobuhiko \textsc{Kusakabe},\altaffilmark{1}\\
Jungmi \textsc{Kwon},\altaffilmark{5}
Taro \textsc{Matsuo},\altaffilmark{15}
Satoshi \textsc{Mayama},\altaffilmark{5}
Michael W. \textsc{McElwain},\altaffilmark{13}
Shoken \textsc{Miyama},\altaffilmark{16}\\
Jun-Ichi \textsc{Morino},\altaffilmark{1}
Amaya \textsc{Moro-Martin},\altaffilmark{17}
Tetsuo \textsc{Nishimura},\altaffilmark{6}
Tae-Soo \textsc{Pyo},\altaffilmark{6}
Eugene \textsc{Serabyn},\altaffilmark{18}\\
Hiroshi \textsc{Suto},\altaffilmark{1}
Michihiro \textsc{Takami},\altaffilmark{19}
Naruhisa \textsc{Takato},\altaffilmark{6}
Hiroshi \textsc{Terada},\altaffilmark{6}
Christian \textsc{Thalmann},\altaffilmark{20}\\
Daigo \textsc{Tomono},\altaffilmark{6}
Edwin L. \textsc{Turner},\altaffilmark{10,21}
Makoto \textsc{Watanabe},\altaffilmark{22}
John P. \textsc{Wisniewski},\altaffilmark{23}\\
Toru \textsc{Yamada},\altaffilmark{24}
Hideki \textsc{Takami},\altaffilmark{1}
Tomonori \textsc{Usuda},\altaffilmark{6}
Motohide \textsc{Tamura}\altaffilmark{1}
}

\altaffiltext{1}{
National Astronomical Observatory of Japan, 2-21-1 Osawa,
Mitaka, Tokyo, 181-8588, Japan
}

\altaffiltext{2}{
Department of Astronomy,
The University of Tokyo, 7-3-1 Hongo, Bunkyo-ku, Tokyo,
113-0033, Japan
}

\altaffiltext{3}{
Department of Earth and Planetary Science,
The University of Tokyo, 7-3-1 Hongo, Bunkyo-ku, Tokyo,
113-0033, Japan
}

\altaffiltext{4}{
Department of Physics,
The University of Tokyo, 7-3-1 Hongo, Bunkyo-ku, Tokyo,
113-0033, Japan
}

\altaffiltext{5}{
The Graduate University for Advanced Studies,
2-21-1 Osawa, Mitaka, Tokyo 181-8588, Japan
}

\altaffiltext{6}{
Subaru Telescope, 650 North A'ohoku Place, Hilo, HI 96720, USA
}

\altaffiltext{7}{
Department of Earth and Planetary Sciences, Tokyo Institute of Technology,\\
Ookayama, Meguro-ku, Tokyo 152-8551, Japan
}

\altaffiltext{8}{
Laboratoire Lagrange (UMR 7293), Universit\'{e} de Nice-Sophia
Antipolis, CNRS,\\
Observatoire de la C\^{o}te d'Azur, 28 avenue Valrose,
06108 Nice Cedex 2, France
}

\altaffiltext{9}{
Max Planck Institute for Astronomy, K\"{o}nigstuhl 17, 69117
Heidelberg, Germany
}

\altaffiltext{10}{
Department of Astrophysical Sciences, Princeton University,
Peyton Hall, Ivy Lane, Princeton, NJ 08544, USA
}

\altaffiltext{11}{
Department of Physics and Astronomy,
College of Charleston, 58
Coming St., Charleston, SC 29424, USA
}

\altaffiltext{12}{
Universit\"{a}ts-Sternwarte M\"{u}nchen,
Ludwig-Maximilians-Universit\"{a}t,
Scheinerstr. 1, 81679 M\"{u}nchen, Germany
}

\altaffiltext{13}{
Exoplanets and Stellar Astrophysics Laboratory,
Code 667, Goddard
Space Flight Center, Greenbelt, MD 20771 USA
}

\altaffiltext{14}{
Institute for Astronomy, University of Hawaii, 640 N. A'ohoku
Place, Hilo, HI 96720, USA
}

\altaffiltext{15}{
Department of Astronomy, Kyoto University,
Kitashirakawa-Oiwake-cho, Sakyo-ku, Kyoto, Kyoto 606-8502, Japan
}

\altaffiltext{16}{
Hiroshima University, 1-3-2, Kagamiyama, Higashihiroshima,
Hiroshima 739-8511, Japan
}

\altaffiltext{17}{
Department of Astrophysics, CAB-CSIC/INTA, 28850 Torrej\'{o}n de Ardoz,
Madrid, Spain
}

\altaffiltext{18}{
Jet Propulsion Laboratory, California Institute of Technology,
Pasadena, CA, 171-113, USA
}

\altaffiltext{19}{
Institute of Astronomy and Astrophysics, Academia Sinica, P.O. Box
23-141, Taipei 10617, Taiwan
}

\altaffiltext{20}{
Astronomical Institute "Anton Pannekoek", University of Amsterdam,\\
Postbus 94249, 1090 GE, Amsterdam, The Netherlands
}

\altaffiltext{21}{
Kavli Institute for Physics and Mathematics of the Universe,\\
The University of Tokyo, 5-1-5, Kashiwanoha, Kashiwa,
Chiba 277-8568, Japan
}

\altaffiltext{22}{
Department of Cosmosciences, Hokkaido University,
Kita-ku, Sapporo,
Hokkaido 060-0810, Japan
}

\altaffiltext{23}{
H.L Dodge Department of Physics \& Astronomy, University of Oklahoma,
440 W Brooks St Norman, OK 73019, USA
}

\altaffiltext{24}{
Astronomical Institute, Tohoku University, Aoba-ku, Sendai, Miyagi
980-8578, Japan
}

\altaffiltext{25}{
NAOJ Fellow
}
\email{norio.narita@nao.ac.jp}

%% `\KeyWords{}' always has to be placed before `\maketitle'.
%Do NOT move this preamble from here!

\maketitle

\begin{abstract}
We report that HAT-P-7 has a common proper motion stellar companion.
The companion is located at $\sim3.9$ arcsec to the east
and estimated as an M5.5V dwarf based on its colors.
We also confirm the presence of the third companion,
which was first reported by \citet{2009ApJ...703L..99W},
based on long-term radial velocity measurements.
We revisit the migration mechanism of HAT-P-7b
given the presence of those companions,
and propose sequential Kozai migration
as a likely scenario in this system.
This scenario may explain the reason for an outlier in the discussion
of the spin-orbit alignment timescale for HAT-P-7b
by \citet{2012ApJ...757...18A}.
\end{abstract}

\section{Introduction}

To uncover formation mechanisms of diverse exoplanetary systems,
the (mis)alignment between the planetary orbital axis and the stellar spin axis,
which can be measured via the Rossiter-McLaughlin (RM) effect
(e.g., \cite{2005ApJ...622.1118O,2011ApJ...742...69H})
or spot-crossing events (e.g., \cite{2011ApJ...743...61S,2012Natur.487..449S}),
has been recognized as an useful clue.
Previous observations have revealed that about one third of hot Jupiters
have tilted or even retrograde orbits relative to their host star's spin
(e.g., \cite{2008A&A...488..763H, 2009PASJ...61L..35N,2009ApJ...703L..99W}). 
This indicates that not only disk-planet interaction but also
few-body interaction (planet-planet scattering and/or the Kozai mechanism)
play an important role in planetary migration processes.
Moreover, some studies pointed out the interesting facts that
spin-orbit misalignments in planetary orbits are apparently correlated
with host star's temperature \citep{2010ApJ...718L.145W} and
age \citep{2011A&A...534L...6T},
while these correlations can be explained
by the properties and evolution of
the internal structure in host stars \citep{2012ApJ...757...18A}.
Those correlations are important clues to understand
the whole picture of planetary migration.

However, previous discussions often overlooked
the possible presence of faint distant companions.
In most of the RM measurements, except for some obvious cases
(e.g., XO-2, HD80606), observers did not check whether
the host star has outer companions.
Thus any correlation between the spin-orbit misalignment and
the existence of binary companions has not been well investigated.
To solve this problem, we have started high-contrast direct imaging
observations for known transiting planetary systems in the course of
the SEEDS (Strategic Explorations of Exoplanets and Disks with Subaru;
\cite{2009AIPC.1158...11T}) project.

\citet{2010PASJ...62..779N} reported two candidate companions around HAT-P-7,
which was known to have a retrograde hot Jupiter HAT-P-7b
\citep{2009PASJ...61L..35N,2009ApJ...703L..99W,2012ApJ...757...18A}.
In this letter, we present evidence that one of the two candidate companions
is indeed a true companion of HAT-P-7, confirmed by the common proper motion
and the distance from Earth inferred from its spectral type
and apparent magnitude.

%%%%%%%%%%%%%%%%%%%%%%%%%%%%%%%%%%%%%%%%%%%%%%%%%%%%%%%%%%%%%%%%%%%%%%
\begin{figure}[thb]
 \begin{center}
  \FigureFile(85mm,85mm){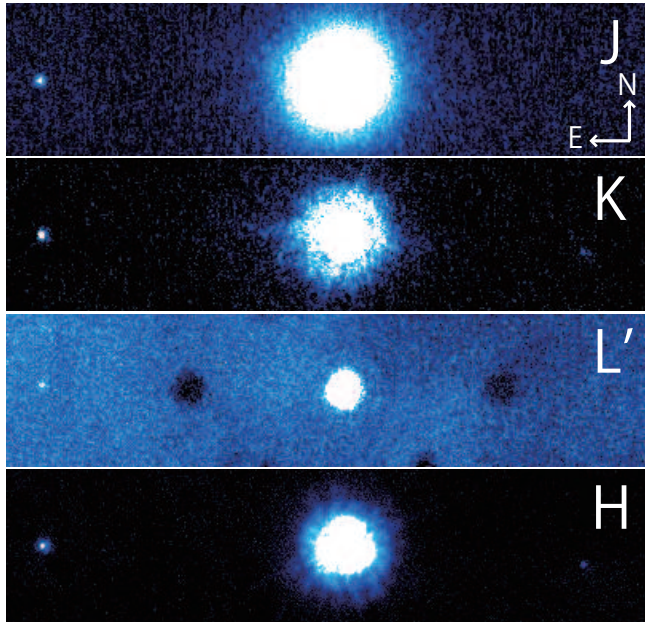}\vspace{-2mm}
 \end{center}
  \caption{High-contrast images around HAT-P-7 taken with the Subaru
  IRCS ($J, K, L^{\prime}$) in 2011 and HiCIAO ($H$) in 2012 (see table~1).
  The field of view of each panel is 8 arcsec (horizontal) $\times$ 2 arcsec (vertical).
  Black circle regions in the $L^{\prime}$ band image are vestiges of dithering pattern.
  North is up and east is left for all panels. Note that only unsaturated images
  are used in this figure to highlight the eastern companion. The western companion
  is not detected in the $J$ and $L^{\prime}$ band images.}
\end{figure}
%%%%%%%%%%%%%%%%%%%%%%%%%%%%%%%%%%%%%%%%%%%%%%%%%%%%%%%%%%%%%%%%%%%%%%

%%%%%%%%%%%%%%%%%%%%%%%%%%%%%%%%%%%%%%%%%%%%%%%%%%%%%%%%%%%%%%%%%%%%%%
\begin{figure}[thb]
 \begin{center}
  \FigureFile(85mm,85mm){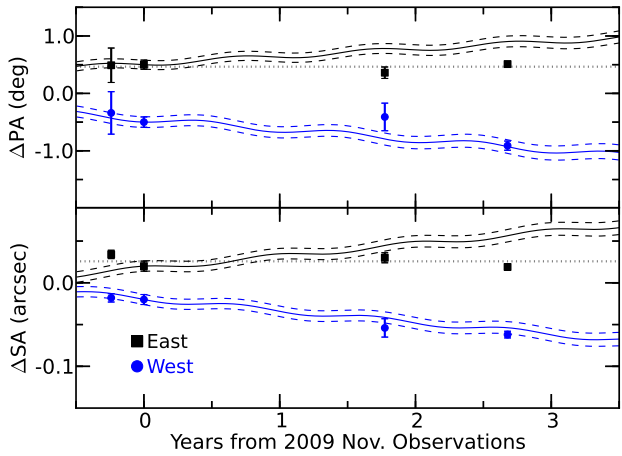}\vspace{-2mm}
 \end{center}
  \caption{
  The panels illustrate the variations of position angles (top) and separation
  angles (bottom) of the both candidates over the course of three years.
  The solid trajectories show the variations expected from an assumption
  that the candidates are background stars: the dashed lines show the 1$\sigma$
  errors. Note that we add offsets to all the values for clarify.
  The dotted lines indicate the averages of observed separation angles
  and position angles for the east candidate. As shown in this figure,
  the east one is inconsistent with a background source, while the west
  object is in good agreement with the background hypothesis.
  }
\end{figure}
%%%%%%%%%%%%%%%%%%%%%%%%%%%%%%%%%%%%%%%%%%%%%%%%%%%%%%%%%%%%%%%%%%%%%%

%%%%%%%%%%%%%%%%%%%%%%%%%%%%%%%%%%%%%%%%%%%%%%%%%%%%%%%%%%%%%%%%%%%%%%
\begin{table*}[thb]
\begin{center}
\caption{Observation logs and magnitudes of candidate companions}
\begin{tabular}{ccccccc}
\hline
UT Dates & Inst. & Band & Exp. Time [s] & Mag.(East) & Mag.(West) & Ref.\\
\hline
2009Aug06  & HiCIAO & $H$ & 585$^{a}$ & $15.12\pm0.04$ & $16.64\pm0.06^{b}$  & this work$^{b}$ \\
2009Oct30  & AstraLux & $i^{\prime}$ & 30 & $18.50\pm0.21$ &  non-detection & \citet{2010PASJ...62..779N}\\
2009Oct30  & AstraLux & $z^{\prime}$ & 30 & $17.43\pm0.09$ &   non-detection & \citet{2010PASJ...62..779N}\\
2009Nov02  & HiCIAO & $H$ & 100$^{c}$ & $15.07\pm0.04$ & $16.56\pm0.07$  & this work \\
2011Aug12  & IRCS & $J$ & 560.6$^{d}$ & $15.81\pm0.19$ & $17.48\pm0.25$ & this work\\
2011Aug12  & IRCS & $K$ & 560.6$^{d}$ & $14.81\pm0.07$ & $16.38\pm0.09$  & this work\\
2011Aug12  & IRCS & $L^{\prime}$ & 396$^{e}$ & $14.45\pm0.41$ &  non-detection &this work\\
2012Jul07  & HiCIAO & $H$ & 1810$^{f}$ & $15.18\pm0.07$  & $16.66\pm0.07$ & this work\\
\hline
\multicolumn{7}{l}{\hbox to 0pt{\parbox{155mm}{\footnotesize
$^{a}$ 19.5 s (saturated) $\times$ 30 images.
$^{b}$ The previous value in \citet{2010PASJ...62..779N} is turned out to be false, and here we present
the corrected value.
$^{c}$ 4.18 s (unsaturated) $\times$ 24 images.
$^{d}$ 180 s (saturated) $\times$ 3 images and 4.12 s (unsaturated) $\times$ 5 images.
$^{e}$ 12 s (unsaturated) $\times$ 33 images.
$^{f}$ 20 s (saturated) $\times$ 89 images and 1.5 s (unsaturated) $\times$ 20 images.
}\hss}}
\end{tabular}
%\end{center}
%\end{table*}
%%%%%%%%%%%%%%%%%%%%%%%%%%%%%%%%%%%%%%%%%%%%%%%%%%%%%%%%%%%%%%%%%%%%%%
%%%%%%%%%%%%%%%%%%%%%%%%%%%%%%%%%%%%%%%%%%%%%%%%%%%%%%%%%%%%%%%%%%%%%%
%\begin{table*}[thb]
%\begin{center}
\caption{Separation angles (SA) and position angles (PA) of candidate companions}
\begin{tabular}{ccccccc}
\hline
UT Dates & Inst. &  SA(East) [''] & PA(East) [$^{\circ}$] & SA(West) [''] & PA(West) [$^{\circ}$] & Ref. \\
\hline
2009Aug06  & HiCIAO & $3.875\pm0.005$ & $89.81\pm0.30$ & $3.139\pm0.005$  & $266.30\pm0.37$  & \citet{2010PASJ...62..779N} \\
2009Nov02  & HiCIAO & $3.861\pm0.006$  & $89.82\pm0.08$ & $3.137\pm0.006$  & $266.14\pm0.09$  & this work\\
2011Aug12  & IRCS &  $3.871\pm0.006$ & $89.68\pm0.10$ & $3.103\pm0.011$ & $266.23\pm0.24$  & this work\\
2012Jul07  & HiCIAO & $3.860\pm0.004$  & $89.83\pm0.05$ & $3.095\pm0.004$  & $265.73\pm0.08$  & this work\\
\hline
\multicolumn{7}{l}{\hbox to 0pt{\parbox{155mm}{\footnotesize
}\hss}}
\end{tabular}
\end{center}
\end{table*}
%%%%%%%%%%%%%%%%%%%%%%%%%%%%%%%%%%%%%%%%%%%%%%%%%%%%%%%%%%%%%%%%%%%%%%

\section{Observations and Results}

For high-contrast direct imaging, we employed
IRCS (for $J$, $K$, $L^{\prime}$ bands) 
and HiCIAO (for $H$ band) with AO188 \citep{2008SPIE.7015E..25H}
on the Subaru 8.2m telescope.
Our observation logs (observing dates, instruments, filters, exposure times) 
and properties of the candidate companions are summarized in tables~1~and~2.
Magnitudes of the bright companion are derived by relative photometry with
the host star using unsaturated images, and then magnitudes of the faint
companion are similarly derived by relative photometry with
the bright companion using saturated images.
Figure~1 shows pictures of HAT-P-7 in different epochs and bands.
Figure~2 plots time changes of positions of the candidate companions.
Based on the Tycho-2 Catalogue \citep{2000A&A...355L..27H},
HAT-P-7 has a proper motion of $-14.8 \pm 1.5$ mas/yr in right ascension (RA)
and $8.7 \pm 1.4$ mas/yr in declination (Dec).
From figure~2, we find that the motion of the western (fainter) candidate is consistent
with a background star ($\chi^2 = 4.3$ for a degree of freedom of 6),
while the motion of the eastern (brighter) candidate  is different from
a background star ($\chi^2 = 83.0$ for a degree of freedom of 6) and
consistent with a common proper motion with HAT-P-7.

In addition to the common proper motion, the spectral type of the eastern
candidate supports a consistent distance with HAT-P-7, as follows.
The distance to HAT-P-7 was estimated as $320^{+50}_{-40}$ pc
by \citet{2008ApJ...680.1450P}, based on the Yonsei-Yale stellar evolution
models \citep{2001ApJS..136..417Y} and
SME (Spectroscopy Made Easy: \cite{1996A&AS..118..595V}) analysis.
On the other hand, using the available colors
($i^{\prime}$, $z^{\prime}$, $J$, $H$, $K$),
we confirm that the eastern companion is an M5.5V dwarf
(see figure~3).
Based on the absolute magnitude of an M5.5V dwarf in $H$ band
\citep{2007AJ....134.2340K}, we estimate a distance modulus of
$\sim6.15$, which corresponds to $\sim300$ pc.
Thus the spectro-photometric distance to the eastern companion
is in good agreement with the distance to HAT-P-7 from Earth.

We thus conclude that HAT-P-7 has a common proper motion
stellar (M5.5V) companion at a projected separation of $\sim3.9$ arcsec
($1240^{+190}_{-160}$~AU).
Note that assuming a random distribution for the companion orbit,
the expected unprojected separation is $4 / \pi$ times larger than
the projected separation (see also table~6 of \cite{2011ApJ...733..122D}
for a conversion factor from the projected separation to
the semi-major axis).
Here we name the eastern companion as ``HAT-P-7B'' (hereafter, just ``B'').

%%%%%%%%%%%%%%%%%%%%%%%%%%%%%%%%%%%%%%%%%%%%%%%%%%%%%%%%%%%%%%%%%%%%%%
\begin{figure}[thb]
 \begin{center}
  \FigureFile(75mm,75mm){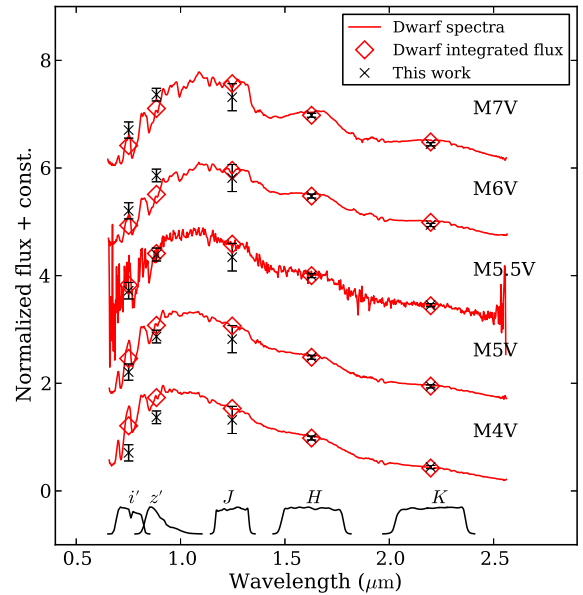}
 \end{center}
  \caption{Comparions of the observed colors with template spectra of
  known M4V-M7V dwarfs \citep{2004AJ....127.2856B,2008ApJ...681..579B}
  taken with IRTF/SpeX. We find the M5.5V dwarf template gives a minimum $\chi^2$.}
\end{figure}
%%%%%%%%%%%%%%%%%%%%%%%%%%%%%%%%%%%%%%%%%%%%%%%%%%%%%%%%%%%%%%%%%%%%%%

%%%%%%%%%%%%%%%%%%%%%%%%%%%%%%%%%%%%%%%%%%%%%%%%%%%%%%%%%%%%%%%%%%%%%%
\begin{figure}[thb]
 \begin{center}
  \FigureFile(85mm,85mm){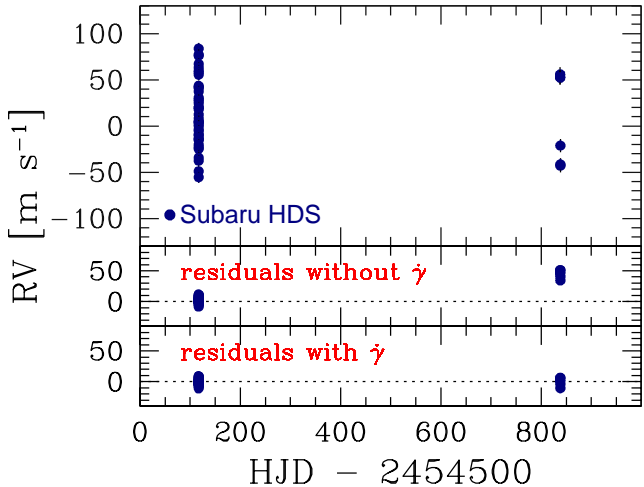}\vspace{-10mm}
 \end{center}
  \caption{Top panel: RVs of HAT-P-7 observed with the Subaru
  HDS. Middel panel: Residuals of RVs from the best-fit one-planet model
  without subtracting the long-term RV trend. Bottom panel:
  Same as the middle panel but with subtracting the RV trend.}
\end{figure}
%%%%%%%%%%%%%%%%%%%%%%%%%%%%%%%%%%%%%%%%%%%%%%%%%%%%%%%%%%%%%%%%%%%%%%

\section{Discussions}

\subsection{Further Evidence of the Third Companion}

\citet{2009ApJ...703L..99W} reported the possible existence of a third
(different from HAT-P-7b and HAT-P-7B) companion from
a long-term radial velocity (RV) trend observed with
the Keck HIRES between 2007 and 2009.
We confirm the long-term trend based on RVs measured with
the Subaru HDS in 2008 and 2010
(the RVs are available upon request).
Figure~4 shows the RVs taken with the Subaru HDS and 
residuals from the best-fit one-planet model with and without
a long-term linear RV trend $\dot{\gamma}$
(see \cite{2011PASJ...63L..67N}, for the fitting procedure).
We find $\dot{\gamma} = 20.3\pm1.8 $~m~s$^{-1}$~yr$^{-1}$,
which agrees well with \citet{2009ApJ...703L..99W}
($\dot{\gamma} = 21.5\pm2.6 $~m~s$^{-1}$~yr$^{-1}$).
Note that ``B'' cannot explain the observed long-term RV trend.
Thus we conclude that there is the third companion in this system.
Here we name the long-term RV companion as ``HAT-P-7c'' (hereafter, just ``c'').

The trend corresponds to the mass constraint of ``c'' as,
\begin{equation}
M_{\rm c} \sin i_{\rm c} / a_{\rm c}^2 \sim 0.12 \pm 0.01\,\,\,
M_{\rm Jup}\,\,{\rm AU}^{-2}.
\end{equation}
The orbital period of ``c'' is presumably longer than 10~yr,
since the trend is still approximately linear.
Thus ``c'' is likely to be more massive than Jupiter.
We do not detect ``c'' in high-contrast direct images,
because it is too close to the host star.

\subsection{Migration Mechanism of HAT-P-7b Revisited}

\citet{2010PASJ...62..779N} pointed out that HAT-P-7b (hereafter, just ``b'')
cannot be migrated by the Kozai mechanism caused by ``B''
in the presence of ``c'' in the system.
However, they also pointed out that there is another possibility of ``sequential''
Kozai migration (see e.g., \cite{2008ApJ...683.1063T,2010AsBio..10..733K})
in this system.
Namely, an inclined outer stellar companion ``B'' causes the Kozai mechanism
for the outer ``c'' first, and then the inclined ``c'' induces the Kozai mechanism
for the inner ``b''.
Such an initial inclined configuration can be formed by
planet-planet scattering of ``b'' and ``c''
(and other possible undetected massive planets).

Recently, \citet{2012ApJ...757...18A} pointed out the interesting fact
that the spin-orbit alignment timescale of HAT-P-7 due to the tidal effect
is much shorter than the star's age
(see figure~25 of \cite{2012ApJ...757...18A}).
Namely, the retrograde orbit of HAT-P-7b appears to be
a strong outlier which should be aligned at the age of HAT-P-7.
However, they did not consider the existence of
the outer companions ``B'' and ``c''.
Here we propose an alternative reason of the outlier:
if the Kozai oscillation of ``b'' caused by ``c''
had once happened,
the tidal timescale for spin-orbit alignment can be much longer.
Based on equation (1) and the timescale of the Kozai oscillation as,
\begin{equation}
P_{\rm Kozai} \sim \frac{M_s}{M_{\rm c}} \frac{P_{\rm c}^2}{P_{\rm b}} (1-e_{\rm c}^2)^{3/2},
\end{equation}
where $M_{\rm s}$ is the host star's mass,
$M_{\rm c}$, $P_{\rm c}$, and $e_{\rm c}$ are the mass, orbital period, and eccentricity of ``c'',
and $P_{\rm b}$ is the orbital period of ``b'' \citep{2007ApJ...670..820W},
we estimate that the timescale of Kozai oscillation of ``b'' caused by ``c''
is comparable to the age of HAT-P-7 ($2.14 \pm 0.26$~Gyr).
Thus a possible slowly-changing orbital inclination due to the Kozai oscillation
might have prevented
HAT-P-7b from achieving a spin-orbit alignment.
We note that, however, the above discussion may be still optimistic,
as the Kozai oscillation is easily suppressed by other effects such as
general relativity, tides, stellar distortion, or extra bodies
(see e.g., \cite{2007ApJ...669.1298F}).
At the current position of ``b'', the timescale of general relativity is indeed
shorter than the timescale of Kozai oscillation of ``b'' caused by ``c''.
This means the Kozai oscillation does not extend the timescale
of spin-orbit alignment any more, although it might have once happened.
We thus note that we may still overlook other special conditions that inhibit
the spin-orbit alignment in this system.

\subsection{Suggestion to Further Discussions on Planetary Migration}

Thus far, the existence of possible faint outer companions around
planetary systems has not been checked and is often overlooked,
even though the Kozai migration models assume the presence of
an outer companion.
To further discuss planetary migration using the information of the RM effect /
spot-crossing events as well as significant orbital eccentricities,
it is important to incorporate information regarding
the possible or known existence of binary companions.
This is also because a large fraction of the stars in the universe form
binary systems \citep{1993AJ....106.2005G}.
Thus it would be important to check the presence of faint binary companions
by high-contrast direct imaging.
In addition, if any outer binary companion is found, it is also necessary to
consider the possibility of sequential Kozai migration in the system,
since planet-planet scattering, if it occurs, is likely to form the initial condition
of such planetary migration.

\section{Summary}

We present evidence that HAT-P-7 has a common proper motion
stellar companion by high-contrast direct imaging with
Subaru HiCIAO and IRCS.
The companion is located at $\sim3.9$ arcsec to the east
and estimated as an M5.5V star based on its colors.
We also confirm the presence of the third companion HAT-P-7c
by RV measurements with the Subaru HDS.

Our finding that HAT-P-7 is a wide binary system increase
the potential that the sequential Kozai migration had once occured
in this system.
This scenario may be
favorable to explain the discrepancy between
the spin-orbit alignment timescale for HAT-P-7b and the stellar age
noted by \citet{2012ApJ...757...18A}.

In recent years, a number of RM and spot-crossing observations have provided
useful information on the spin-orbit (mis)alignment, which is an important
clue to understand the variety of planetary migration mechanisms.
The same holds for eccentric planetary systems,
since the eccentricity is another important clue to
discuss planetary migration by few-body interactions.
We thus propose that high-contrast direct imaging
observations should be made for known planetary systems
to check the presence of outer faint companions.
Since the dependence of the spin-orbit (mis)alignment
or significant eccentricity on the existence of
outer companions is still unclear,
it is important to conduct such observations
to understand the entire picture of planetary migration.

\bigskip

This paper is based on data collected at Subaru Telescope,
which is operated by the National Astronomical Observatory of Japan (NAOJ).
This research has benefitted from the SpeX Prism Spectral Libraries,
maintained by Adam Burgasser at http://pono.ucsd.edu/~adam/browndwarfs/spexprism.
We are grateful to the referee, Dr. Simon Albrecht, for insightful suggestions.
N.N. acknowledges supports by NAOJ Fellowship,
NINS Program for Cross-Disciplinary Study,
and the Japan Society for Promotion of Science (JSPS)
Grant-in-Aid for Research Activity Start-up No. 23840046.
This work is partly supported by JSPS Fellowships for Research
(DC1: 22-5935, 23-271),
a Grant-in-Aid for Specially Promoted Research, No. 22000005 from
the Ministry of Education, Culture, Sports, Science and Technology (MEXT),
grant AST-1009203 from the National Science Foundation,
and by the Mitsubishi Foundation.
Part of this research was carried out at the Jet Propulsion Laboratory,
California Institute of Technology, under a contract with
the National Aeronautics and Space Administration.

%%%%%%%%%%%%%%%%%%%%%%%%%%%%%%%%%%%%%%%%%%%%%%%%%%%%%%%%%%%%%%%%%%%%%%

%%%%%%%%%%%%%%%%%%%%%%%%%%%%%%%%%%%%%%%%%%%%%%%%%%%%%%%%%%%%%%%%%%%%%%

\end{document}